\documentclass[fleqn,10pt]{wlscirep}
\usepackage[utf8]{inputenc}
\usepackage[T1]{fontenc}
\usepackage{graphicx}%
\usepackage{multirow}%
\usepackage{amsmath,amssymb,amsfonts}%
\usepackage{amsthm}%
\usepackage{mathrsfs}%
\usepackage[title]{appendix}%
\usepackage{xcolor}%
\usepackage{float}
\usepackage{textcomp}%
\usepackage{manyfoot}%
\usepackage{booktabs}%
\usepackage{algorithm}%
\usepackage{algorithmicx}%
\usepackage{algpseudocode}%
\usepackage{listings}%
\usepackage{subcaption}
\usepackage[most]{tcolorbox}
\newenvironment{llmbox}{}{}
\tcolorboxenvironment{llmbox}{
  boxrule=0.5pt,
  colback=gray!5,
  colframe=gray!50!black,
  arc=2mm,
  left=2mm,
  right=2mm,
  top=1mm,
  bottom=1mm,
  enhanced,
  breakable
}
\title{LLM-Enabled NWDAF: A Step Toward AI-Native 6G Network Intelligence}

\author[1]{Henok Daniel}
\author[1,*]{Omar Alhussein}
\author[2]{Cheng Li}
\author[2]{Jie Liang}
\author[3,4]{Ernesto Damiani}
\affil[1]{KU 6G Research Center, College of Computing and Mathematical Sciences, Khalifa University, Abu Dhabi, UAE}
\affil[2]{School of Engineering Science, Simon Fraser University, Vancouver, Canada}
\affil[3]{Center of Cyber-Physical Systems (C2PS), Department of Computer Science, Khalifa University, Abu Dhabi, UAE}
\affil[4]{Department of Computer Science, Università degli Studi di Milano, Milano, Italy}

\affil[*]{omar.alhussein@ku.ac.ae}

\begin{abstract}
The Network Data Analytics Function (NWDAF) is central to enabling zero-touch network management in fifth-generation (5G) networks by supporting real-time analytics and closed-loop automation. Despite its critical role, open-source NWDAF implementations remain limited in scope and accessibility. In this paper, we develop an open-source NWDAF, compatible with the open-source core network Free5GC, that collects network data via subscriptions to Network Functions (NFs), and also includes an integrated Large Language Model (LLM) interface that enables natural language interaction with human operators. The interface processes user intents, encodes them using a semantic embedding model, and maps them to one of seven predefined intent categories to trigger analytics queries or event subscription commands. This architecture abstracts the complexity of traditional interfaces, allowing non-expert users to manage network analytics and subscriptions with ease. The system supports Access and Management Function (AMF) and Session Management Function (SMF) event subscriptions, real-time monitoring, and analytics retrieval via Prometheus, all accessible through a conversational interface. By bridging AI-driven intent recognition with standardized network analytics, our implementation enhances operator usability and provides a foundation towards AI-native 6G networks. The source code and datasets generated during the current study are available in the github repository, https://github.com/HenokDanielbfg/testbed.
\end{abstract}
\begin{document}

\flushbottom
% \maketitle
\thispagestyle{empty}

% \noindent Please note: Abbreviations should be introduced at the first mention in the main text – no abbreviations lists. Suggested structure of main text (not enforced) is provided below.

% \keywords{5G core network, 6G, LLM interface, NWDAF, open-source testbed}



\maketitle
\begin{flushleft}
\end{flushleft}
\section*{Introduction}

In the rapidly evolving landscape of telecommunications, the integration of open-source software has revolutionized the development and deployment of 5G testbeds, paving the way for innovative research and practical experimentation in next-generation wireless networks. 5G networks adopt a modular and service-oriented architecture known as Service-Based Architecture (SBA), designed to support diverse requirements across eMBB, mMTC, and URLLC \cite{34}, thereby paving the way for intelligent use cases such as smart cities, extended reality (XR) and autonomous systems \cite{49, 35}. One example of a use case is the VICA counseling robot, which leverages 5G/6G features such as network slicing and Multi-access Edge Computing (MEC) to achieve real-time responsiveness for conversational AI services \cite{48}.
Within SBA, NFs operate as independent microservices that can dynamically discover and communicate with each other through standardized interfaces. Open-source projects such as Free5GC, Open5GS, UERANSIM, srsRAN, and OpenAirInterface (OAI) provide flexible environments for experimentation, enabling developers and researchers to explore, customize, and validate 5G capabilities \cite{12,4,5,6,7}. 

To enable data-driven intelligence, the Third Generation Partnership Project (3GPP)-Release 15 introduced the NWDAF, which aggregates and analyzes data from multiple NFs \cite{14,36}. The NWDAF is designed to play a pivotal role in enabling closed-loop automation in 5G by facilitating the exchange of analytics across NFs. This functionality underpins emerging paradigms such as zero-touch network and service management, where the network can respond autonomously to real-time changes \cite{30,37}. Recent studies have demonstrated NWDAF’s potential to support life-cycle automation and predictive control in slicing and orchestration scenarios \cite{40}.

At the same time, there is increasing demand for tools that make these systems more accessible, both for operators and orchestrators, by abstracting away low-level control in favor of higher-level intent. This vision is embodied in the emerging paradigm of Intent-Based Networking (IBN), which seeks to enable operators to express desired outcomes rather than execution steps, enabling high-level declarative control that is translated into deployable configurations through intent translation systems \cite{30, 43}. The authors in \cite{39} have compiled a document that explains the concepts and definitions of IBN. This approach has already been prototyped in private 5G networks using natural language processing to simplify configuration and orchestration \cite{32}.

IBN represents a shift from procedural to declarative network management. Instead of adjusting configurations through Command Line Interfaces (CLI) or Application Programming Interface (API) calls, operators can define objectives such as “ensure low latency for this application” or “monitor registration failures in region X,” leaving the system to resolve these intents through underlying analytics and control loops. This paradigm aligns naturally with the SBA of 5G. However, despite this architectural flexibility, open-source 5G networks currently lack native interfaces to support IBN functionality.

Moreover, as the industry progresses toward 6G, the ability to manage immersive and massive communication workloads will become central in enabling Metaverse applications. These use cases demand not only high bandwidth and ultra-low latency but also intelligent orchestration frameworks capable of adapting to dynamic user behavior and heterogeneous network environments \cite{50}. Semantic interfaces and AI-driven automation, such as those provided by LLM-augmented NWDAF, offer a critical foundation for simplifying management in these emerging systems \cite{51}. Combining natural language interaction with real-time analytics supports the adaptive infrastructure needed for 6G Metaverse environments.

To address the above challenges, we introduce an LLM-powered intent interface built on top of the NWDAF. In most open-source platforms, NWDAF implementations are limited to metrics exposure and lack the user-facing flexibility required to translate human goals into analytics-driven actions. Our NWDAF collects data directly from the core through NF event subscriptions, capturing UE registration/deregistration, session setup, and handover events in real time. These event-driven datasets form the foundation for analytics retrieval and prediction. Additionally, our proposed system integrates an LLM interface into the NWDAF, allowing human-centric network control using natural language. This enables analytical queries such as “show me handover frequency by SUPI” or control intents such as “unsubscribe from AMF events,” which are parsed, interpreted, and executed in real time. The interface acts as a translator between human-readable intents and NWDAF-compatible API calls, making analytics retrieval and event subscription management accessible to users without requiring domain-specific scripting knowledge. Similar efforts have shown that LLMs can accurately translate high-level network queries into actionable tasks in core network environments \cite{41, 42}.

At the core of our LLM interface is an intent classification and translation module. Our system follows a Retrieval-Augmented Generation (RAG) paradigm, where user prompts are semantically matched against a repository of predefined intents before being executed as analytics queries or subscription commands. Once classified, the system communicates with the appropriate internal representation (e.g., REST API calls) to fulfill the intent. This modular design allows the interface to remain extensible; new intent classes or analytics types can be incorporated without overhauling the system architecture \cite{52}.

This integration of LLMs into NWDAF not only advances research in IBN for 5G, but also opens the door to more human-in-the-loop experimentation frameworks where real-time network analytics can be explored through natural interaction. The system lowers barriers for researchers, students, and network operators who wish to explore 5G behaviors without writing custom code or understanding the intricacies of NF interfaces. As the 5G landscape moves toward 6G and beyond, such interfaces will likely become a necessary part of cognitive, adaptive network infrastructure \cite{44}. Our contributions can be summarized as follows:
\begin{itemize}
    \item We extend Free5GC with a fully functional NWDAF that supports AMF/SMF event subscription and real-time notification handling.
    \item We integrate an LLM-powered intent interface to the NWDAF for real-time interaction, enabling both analytics queries and subscription management.
    \item We generate rich network event datasets from AMF and SMF via NWDAF subscriptions, using a custom mobility model in UERANSIM to simulate dynamic UE behavior and trigger mobility-related events.
    \item We evaluate the LLM interface’s intent classification capabilities, achieving 98.43\% accuracy across seven operational categories.
    \item We release our codebase as open-source to support future research on NWDAF capabilities in open 5G testbeds.

\end{itemize}
The remainder of this paper is organized as follows: Section \ref{sec-3} reviews existing open-source 5G projects and related work. Section \ref{sec-4} outlines the system architecture of 5G networks, describing the core, RAN, and UE components. Section \ref{sec-5} presents implementation details, including data collection, event subscription mechanisms, and the LLM interface. Section \ref{sec-6} discusses test results and network performance evaluation. Finally, Section \ref{sec-7} concludes the paper and highlights future directions in 5G network simulation and analytics.

\section*{Related Work}\label{sec-3}
 
The transition to SBA in 5G core networks and the virtualization of Radio Access Networks (RAN) have driven extensive research in 5G core development, User Equipment (UE) behaviour, and data analytics. This section explores current works in these areas, focusing on 5G core and RAN implementation, NWDAF integration, IBN in 5G, and mobility modeling.

\subsection*{5G Core and RAN Implementation}

Several projects have emerged in the area of open source 5G implementations, each contributing uniquely to the development and testing of 5G technologies. Open5GS is an open-source project designed for building and managing new radio/long-term evolution (NR/LTE) mobile networks. It offers a comprehensive 5GC implementation, adhering to 3GPP Release 17 standards. Open5GS provides flexibility and scalability, making it suitable for both research and commercial deployments \cite{6,11}. UERANSIM  stands out as an open-source 5G UE and RAN simulator, offering simulation capabilities for both UE and gNodeB \cite{7}. SrsRAN is an open-source project offering a complete suite for 4G and 5G RAN and core networks. It includes implementations for eNodeB (4G), gNodeB (5G), and UE, along with a core network for LTE \cite{4}. OpenAirInterface (OAI) provides open-source implementations for both 5GC and RAN. The OAI 5GC project aims to deliver a 3GPP-compliant 5G standalone core network with a rich feature set, while the RAN project focuses on implementing the 5G NR interface \cite{5}. IEEE 5G/6G Innovation Testbed is a cloud-based, end-to-end 5G network emulator developed by IEEE to facilitate the testing and experimentation of 5G and emerging 6G technologies \cite{3}. However, although it is composed of open-source components, the project itself is subscription-based. Free5GC is an open-source implementation of the 5GC network based on 3GPP Release 15 and beyond. It supports key functions such as the access and mobility management function (AMF), session management function (SMF), user plane function (UPF), and network repository function (NRF), and adopts a modular SBA \cite{21}. Designed for research and prototyping, Free5GC enables RESTful inter-function communication.

\subsection*{NWDAF Integration}

A federated learning-based architecture for NWDAF is proposed in \cite{15} to mitigate overloading and data privacy issues. Their design features two components: leaf NWDAFs, deployed within each NF to create local models, and a root NWDAF, deployed in the core, which aggregates the locals into a global model. Notably, their work is conceptual, offering an architectural framework and use-case design.

The authors in \cite{2} employ an NWDAF in analyzing a dataset from a simulated 5G network. The analysis particularly focuses on leveraging machine learning models to classify three kinds of network protocols. They tested 2 models (Random Forest and Decision Tree), and achieved an average accuracy of approximately 65\%.

In \cite{10}, the integration of NWDAF with Open5GS and UERANSIM is detailed. This work describes the collection of control-plane signaling data over a 138-minute period by leveraging the 5G service-based architecture to monitor inter-NF communications, such as BSF-NRF interactions, and using these insights to propose optimal placements for core NFs. The resulting dataset is open-sourced and available on GitHub.

In \cite{1}, the authors develop an NWDAF compatible with Free5GC. The NWDAF’s architecture aligns with that of the other Free5GC NFs. Their module comprises a model training logical function (MTLF) for training machine learning models and an analytics logical function (AnLF) for delivering inference results to consumers \cite{22}. They also successfully registered the NWDAF with the NRF. For analytics, they utilize the MNIST dataset, leaving the integration with other NFs for live data collection and analytics as a future step. Their code is available on GitHub, which we have used as a foundation for our current implementation.

\subsection*{Intent-Based Networking in 5G}

The integration of IBN into 5G infrastructures has emerged as a promising approach to achieve autonomous, intelligent control over network behavior. Recent works have explored various architectural and practical dimensions of this integration, advancing the capabilities of closed-loop automation.

A foundational architecture for IBN in 5G is introduced by \cite{31}, who present an Intent Controller capable of translating high-level objectives into enforceable policies. Their system supports self-configuration and optimization via modules for intent translation, policy enforcement, and continuous audit feedback, enabling full closed-loop control. Expanding on usability, \cite{32} propose a natural language processing (NLP)-powered Intent Engine tailored to simplify private 5G management. Their system allows operators to specify intents in natural language, which are parsed into structured policies for tasks such as slice provisioning, user positioning, and edge service orchestration, demonstrating low-latency translation and ease of use for non-expert stakeholders.

To further enhance automation, \cite{33} combine IBN with the 3GPP-defined NWDAF function to realize proactive service assurance. By leveraging AI-driven analytics to forecast traffic and detect anomalies, their hybrid framework enables preemptive slice scaling and resource allocation. This convergence of analytics and intent control illustrates the synergy between IBN and NWDAF. In a complementary direction, \cite{30} present a practical implementation of LLM-powered intent parsing aligned with 3GPP standards. Using ChatGPT and structured prompts, their system classifies user requests into valid 5G core intents and outlines integration strategies for live NWDAF-enabled environments.

Collectively, these studies highlight a clear trend: the fusion of IBN with AI and LLMs is not only enhancing network automation but also democratizing control through intuitive, intent-driven interfaces. Building on these developments, our work unifies LLM-based intent recognition with real-time analytics and subscription orchestration via NWDAF, offering an end-to-end framework for adaptive and intelligent 5G network management.

\subsection*{Mobility Modeling}

Several academic literatures have discussed different simulated mobility models as they pertain to user behavior in a network. Here, we will consider 3 mobility models:

The Real-Life Mobility Model (RLMM) \cite{16} captures the cyclic routines of a typical working individual by defining discrete behavioral states, such as \textit{at-home}, \textit{at-work}, \textit{at-popular-place}, and distinct travel modes for regular and alternative routes. Each mobile user is assigned static locations for home and work, alongside a personalized set of social destinations. The model introduces behavioral divergence between weekdays and weekends, with customizable pause times and commute distances, effectively emulating daily life rhythms.

Building on structured daily behavior, the Working Day Movement Model (WDMM) \cite{17} integrates multiple submodels to simulate a full working day cycle. Nodes transition through scheduled activities beginning with a randomized wake-up time, followed by morning commutes, extended work periods, and optional evening outings. Intra-office dynamics are captured using Pareto-distributed pause times, while community formation emerges through shared public spaces and coordinated after-hours activities. This composite design yields mobility traces that reflect both individual and social tendencies.

In contrast to purely schedule-based models, the Weighted Waypoint Model (WWP) \cite{18} introduces a probabilistic framework grounded in location popularity. Targeted primarily at pedestrian behavior in campus-like environments, it uses a time-aware Markov chain to guide destination selection based on real-time popularity weights. Locations such as classrooms, libraries, and cafeterias are assigned distinct pause-time distributions. Classrooms tend to follow bell curves, while libraries exhibit heavy-tailed patterns, adding temporal depth and heterogeneity to user movement.

Together, these models span a wide spectrum of mobility behaviors, from structured routines and socially driven schedules to probabilistic destination preferences, providing a comprehensive foundation for simulating realistic user dynamics in networked systems.

\section*{5G System Architecture}\label{sec-4}

\subsection*{Core Network}
5G systems adopt a SBA, marking a major departure from the tightly-coupled, hierarchical designs of earlier mobile cores \cite{9}.  In SBA, each NF exposes its capabilities as discrete services over standardized Service-Based Interfaces (SBIs), typically RESTful APIs \cite{19}.  Because communication is stateless and endpoint-agnostic, an NF can discover and invoke any other NF dynamically rather than relying on pre-configured links. 

Central to this arrangement is the Network Repository Function (NRF), which maintains a registry of all active NFs and their service profiles \cite{20}.  When, for example, the AMF must engage the SMF, it queries the NRF, selects an available SMF instance, and establishes a transient connection.  This mechanism enables on-the-fly load balancing, rapid horizontal scaling, and automatic failover across redundant NF instances.

SBA also applies a producer-consumer paradigm: any NF can simultaneously publish a service (act as producer) and subscribe to external services (act as consumer). That decoupling supports incremental feature growth; new services can be introduced without altering existing components, provided they register with the NRF and conform to the common SBI framework.

\subsection*{Radio Access Network (RAN)}
In the 5G New Radio (NR) RAN, the base station (gNodeB) integrates both control-plane and user-plane logic in alignment with 3GPP specifications \cite{27}.  The control plane (CP) manages tasks such as radio-resource control (RRC), mobility, and QoS negotiation, while the user plane (UP) conveys data employing PDCP, RLC, MAC, and PHY layers.

Communication between gNBs and the core utilizes the NG interface, divided into N2 (signaling) which uses the NG Application Protocol (NGAP) for registration, session setup, and mobility control and N3 (user plane) which tunnels user traffic over GTP-U toward the User Plane Function (UPF) \cite{24,28}.
This two-channel split allows independent scaling and optimization of signaling and data transport paths, a foundation for ultra-low-latency and high-throughput use cases.

\subsection*{User Equipment (UE)}
A 5G User Equipment stack comprises of a Non-Access Stratum (NAS) layer that negotiates registration, authentication, and session management with the core \cite{45} and an Access Stratum (AS) layer that implements the radio protocol stack and interacts with the RAN \cite{46}.

UE operation is typically represented by a state machine cycling through \textit{idle}, \textit{connected}, and \textit{inactive} modes. These states govern paging, handover behavior, and power-saving features. To maintain service quality, UEs continuously measure metrics such as signal strength and throughput, reporting them to the network for mobility and scheduling decisions \cite{46}.

In combination, these three layers, SBA-driven core, NR-based RAN, and standards-compliant UE, form the baseline architecture for 5G deployments.

\section*{Implementation}\label{sec-5}

\begin{figure}
    \centering
    \includegraphics[width=1\linewidth]{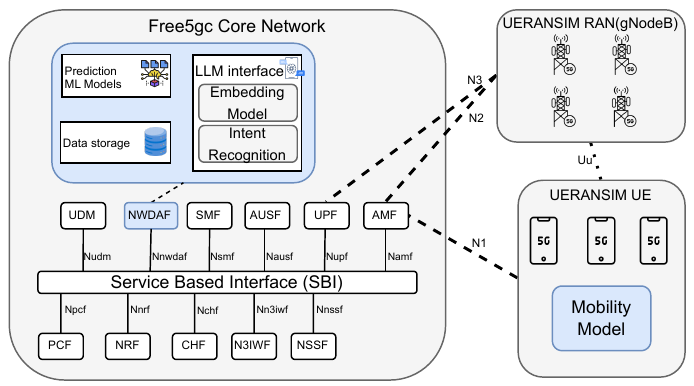}
    \caption{5G Network Architecture with LLM-Powered NWDAF and UERANSIM-based Custom Mobility Model}
    \label{fig:1}
\end{figure}

 Our implementation comprises three main components: the 5G Core Network, Radio Access Network, and User Equipment simulation, as shown in Fig. \ref{fig:1}. Each component is implemented using open-source software and configured to create a comprehensive testing environment. 
 
\subsection*{Core Network Implementation}

The core network is implemented using Free5GC. This platform supports both experimental research and real-world deployments by offering a comprehensive suite of NFs, each responsible for specific tasks within the 5G core\cite{26}. In this architecture, the control plane operations, service discovery, and user plane traffic are efficiently managed through dedicated NFs \cite{21}. Below is a breakdown of these functions:

AMF (Access and Mobility Management Function):
The AMF is responsible for managing initial registration, connection setup, and mobility for UEs \cite{23}. It maintains UE context and tracks location updates to ensure seamless handovers as users move across the network\cite{23}. Additionally, the AMF coordinates with other functions such as the SMF and AUSF to establish secure communication sessions\cite{25}.

SMF (Session Management Function):
The SMF is tasked with establishing, modifying, and terminating data sessions for users\cite{21}. During session setup, it collaborates closely with the AMF to ensure appropriate connection parameters are used. The SMF also configures the UPF to route and manage user data in accordance with session requirements and network slicing strategies\cite{23}.

UPF (User Plane Function):
The UPF handles the forwarding of data packets and ensures low-latency routing for user traffic. It enforces Quality of Service (QoS) policies and makes intelligent routing decisions to enhance network performance. Serving as a bridge between the control and user planes, the UPF interfaces with both radio access networks and external data networks\cite{23}.

NRF (Network Repository Function):
The NRF acts as a centralized service registry, maintaining an updated list of available NFs\cite{26}. By using RESTful APIs, it facilitates efficient service discovery and dynamic communication among NFs\cite{28}. This enables other NFs to locate and interact with appropriate services within the 5G core\cite{22}.

AUSF (Authentication Server Function):
The AUSF is responsible for authenticating users and devices, ensuring that only authorized subscribers gain access to the network. It works in conjunction with the UDM to verify subscriber credentials and establish secure connections\cite{19}. The AUSF plays a pivotal role in initiating the security handshake processes across the 5G network.

CHF (Charging Function):
The CHF monitors data usage and session metrics to generate accurate charging and billing information\cite{28}. It integrates with the PCF to apply real-time charging rules and manage user tariffs effectively. This ensures that users are billed appropriately based on their data consumption.

N3IWF (Non-3GPP Interworking Function):
The N3IWF facilitates connectivity between the 5G core and non-3GPP networks, such as Wi-Fi or fixed broadband. It provides protocol translation and supports seamless handover, maintaining service continuity as users transition between different access networks\cite{21}. Through this capability, the 5G core can extend its reach to a broader range of devices and access technologies.

NSSF (Network Slice Selection Function):
The NSSF determines the most suitable network slice for each UE, based on service demands and subscription profiles\cite{25}. This ensures that users receive the appropriate QoS and resource allocation as defined by their selected network slice. The NSSF supports multi-tenancy and enhances efficiency by adapting network behavior to fit different use cases.

PCF (Policy Control Function):
The PCF makes real-time policy decisions that influence session parameters and overall network behavior. It enforces charging regulations, QoS guidelines, and access restrictions, ensuring optimal use of network resources. Working with the SMF, the PCF can dynamically update configurations based on current network conditions\cite{21}.

UDM (Unified Data Management):
The UDM serves as the central hub for subscriber data, including authentication credentials, user profiles, and service entitlements\cite{26}. It provides vital information to functions such as the AUSF to support secure access. The UDM also ensures consistency of user data across the network, facilitating seamless management\cite{28}.

UDR (Unified Data Repository):
The UDR is a persistent storage solution for subscriber data and network policies. It collaborates with the UDM to maintain high availability and consistency of critical information. In addition to current data, the UDR stores historical session records and other essential information used for analytics and policy enforcement\cite{21}.

Moreover, we have implemented comprehensive event subscription mechanisms within the AMF and SMF. These functions expose RESTful APIs that allow other NFs to subscribe to specific network events. The AMF supports subscriptions for UE registration, deregistration, handover updates, and connectivity state changes. Similarly, the SMF provides subscription capabilities for session management events, including PDU session establishment and modifications. More details on how these subscriptions work can be found in our conference paper \cite{47}.
\begin{figure}
    \centering
    \includegraphics[width=0.5\linewidth]{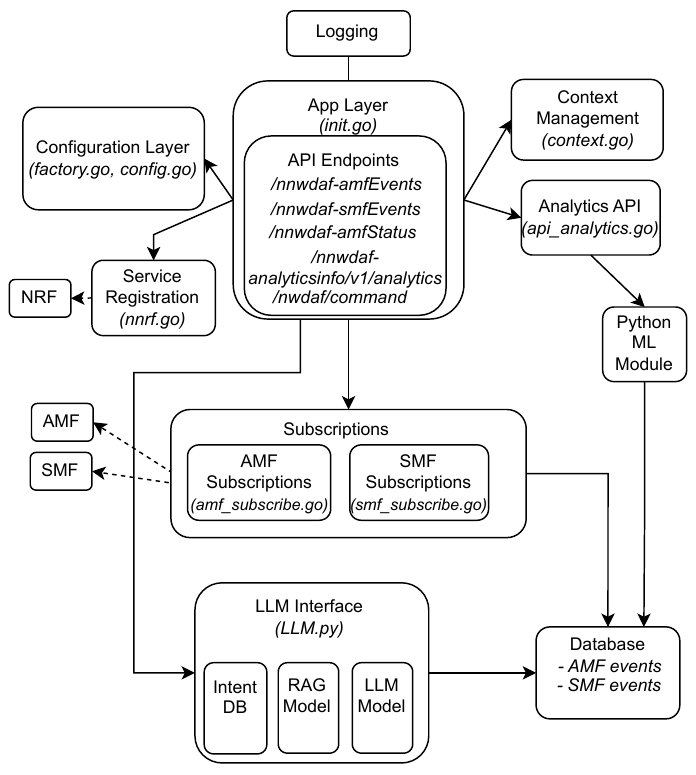}
    \caption{System Architecture of LLM-Enhanced NWDAF with AMF/SMF Subscriptions}
    \label{fig:2}
\end{figure}

% \begin{figure}[]
%     \centering
%     \includegraphics[width=1\linewidth]{Fig3.pdf}
%     \caption{Subscription Mechanism}
%     \label{fig:3}
% \end{figure}
\subsection*{NWDAF}
\begin{figure}[]
    \centering
    \includegraphics[width=0.5\linewidth]{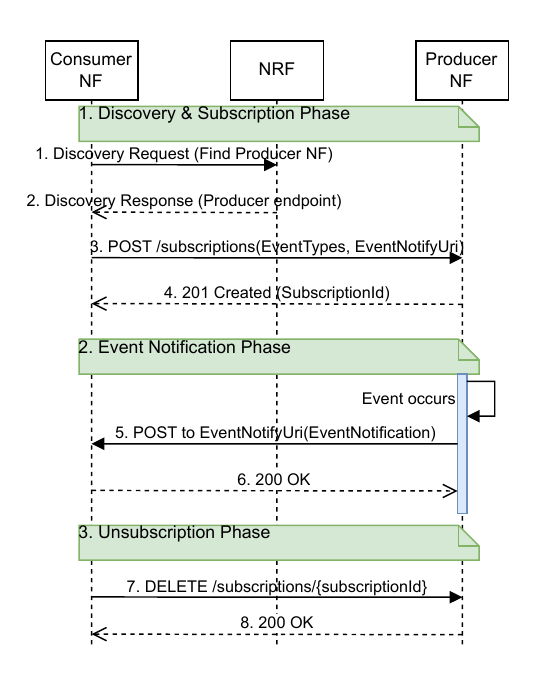}
    \caption{5G Network NF Subscription and Notification Workflow as Defined by 3GPP Standards}
    \label{fig:3}
\end{figure}

The NWDAF architecture is shown in Fig.~\ref{fig:2}. The process of event subscriptions that the NWDAF utilizes is shown in Fig.~\ref{fig:3}. This component collects data from other NFs and processes it to generate network insights to interested parties \cite{8}. Our implementation of the NWDAF builds on the work of \cite{1} who laid out the Free5GC based design of the NF. Our contribution to it includes the addition of the “service” module, data collection mechanism, prediction models, and LLM interface for managing subscriptions and analytical insights. 
The NWDAF integrates with the core by first registering itself with the NRF where it advertises its analytical services\cite{29}. NWDAF reads from an easily configurable YAML file containing a list of events to subscribe to, pertaining to each NF and subscribes to said events via the consumer module.
Our implementation currently supports subscriptions to AMF and SMF events.
% \begin{figure}
%     \centering
%     \includegraphics[width=1\linewidth]{Fig4.pdf}
%     \caption{LLM diagram}
%     \label{fig:4}
% \end{figure}
In addition to its event subscription and notification mechanisms, the NWDAF enhances the core network by integrating predictive analytics that forecast critical network events. By continuously analyzing both historical and real-time data collected from NFs, the NWDAF has the ability to predict key parameters such as the most probable handover cell for user equipment, the expected duration of registration states, and future variations in active user counts. This forecasting layer empowers the network to proactively adjust resource allocation and optimize mobility management, thereby reducing latency and enhancing overall service quality. Currently, we have one prediction model in the core, which predicts the next handover cell for a specific UE.
\begin{figure}
    \centering
    \includegraphics[width=0.5\linewidth]{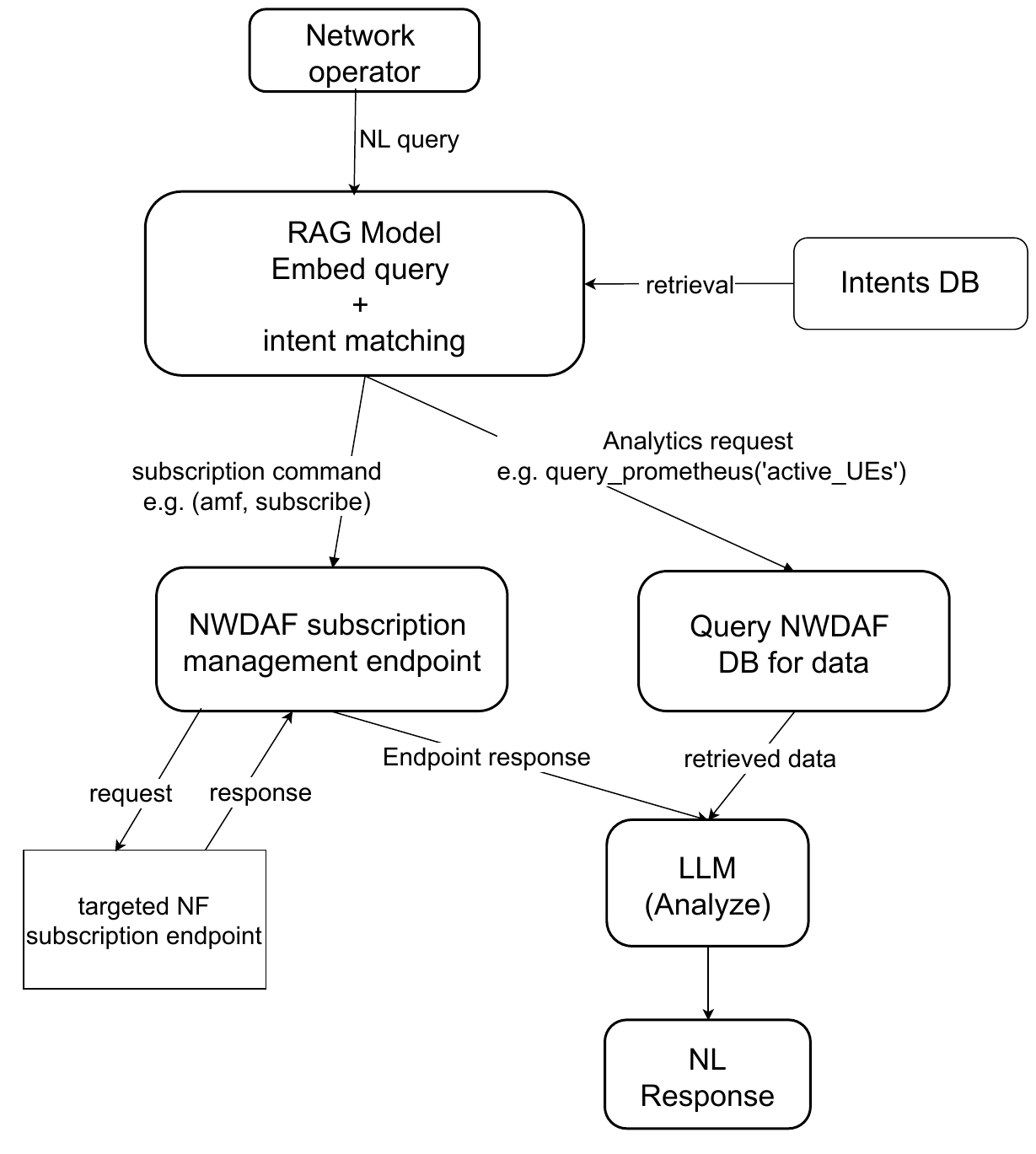}
    \caption{Natural Language Query Processing for NWDAF with RAG Based Intent Matching and LLM Driven Analytics in 5G Core Networks}
    \label{fig:4}
\end{figure}

\subsection*{LLM Interface}
An important addition to the NWDAF framework is the integration of a conversational LLM-based interface, that acts as a middleware between human operators and the NWDAF. As illustrated in Fig.\ref{fig:4}, its core function is to interpret operator intents expressed through natural language and translate them into actionable commands, either querying NWDAF's Prometheus metrics for analytics or managing event subscriptions through NWDAF REST endpoints. The system follows a Retrieval-Augmented Generation (RAG) paradigm to match user input with pre-defined intent categories and then routes the query accordingly.

The system incorporates a flexible mechanism for interpreting operator requests by combining intent embeddings, semantic similarity matching, and function routing. At its foundation, the interface maintains a repository of known intents and their associated categories, stored in a structured text file. These entries represent the different types of requests an operator might issue, ranging from analytical queries such as “How many UEs are active?” to control operations like “Unsubscribe from SMF events.” Each intent is explicitly linked to a category, for instance Analytics Query, AMF Subscribe, or SMF Unsubscribe, which later guides the choice of system function to be invoked.

To support robust and naturalistic interactions, the interface uses an embedding model to represent both the stored intents and the incoming queries as high-dimensional vectors. These embeddings are designed to capture subtle aspects of meaning and linguistic variation, enabling the system to handle operator requests even when phrased differently from the stored reference forms. For example, a query such as “Show me the number of currently active users” would be recognized as equivalent to “How many UEs are active?” despite differences in wording. This ability to abstract beyond surface-level text patterns is essential in real-world environments, where operators often rely on shorthand expressions, colloquial phrasing, or paraphrased instructions.

When an operator issues a query, the system generates its embedding using the same model applied to the intent repository. It then computes the cosine similarity between the query vector and all stored intent vectors. The entry with the highest similarity score is identified as the most relevant intent, and the corresponding category is retrieved. This classification step is critical, as it determines the downstream system behavior and ensures that even ambiguous or loosely phrased queries can be mapped to a precise and actionable instruction.

Once the intent has been classified, the system maps the associated category to a predefined function call. If the identified category corresponds to an analytics operation, the system invokes a Prometheus query function. This function is responsible for constructing a PromQL statement, submitting it to the Prometheus server, and collecting the resulting metric data. The raw response, which may include time series data or counters, is then prepared for higher-level interpretation. On the other hand, if the category corresponds to subscription management, the system instead initiates a control operation by sending a POST request to the NWDAF endpoint. This request contains a JSON payload specifying the desired action, such as subscribing or unsubscribing, along with details of the target network function. In this way, the classification outcome directly determines whether the system operates in an analytical or control capacity.

The final stage of the workflow involves generating a user-facing response that translates raw data or confirmation messages into human-readable summaries. To achieve this, the system integrates the large language model into the response pipeline. The retrieved data, whether from Prometheus or the NWDAF REST API, is provided to the model in JSON format. Alongside this structured input, the message context also includes a system prompt that defines the role of the LLM, the original user query, and the function’s result message. By layering these elements together, the model is able to generate explanations that are not only factually accurate but also contextually appropriate. For instance, when metric data is returned, the LLM may produce a summary such as “Currently, there are 52 active UEs registered across all gNodeBs.” In the case of control operations, the model may respond with a confirmation message such as “Successfully unsubscribed from AMF event notifications.”

Through this design, the intent classification and processing pipeline ensures that operator requests, regardless of phrasing or complexity, can be accurately understood and translated into meaningful system actions. The use of embeddings for semantic similarity, combined with dynamic routing of function calls and natural language response generation, enables a high level of flexibility. It also supports an interaction style that is closer to natural conversation, reducing the need for operators to learn rigid command syntax and allowing them instead to focus on operational objectives.

\subsection*{ RAN \& UE Implementation} 

The RAN is implemented using UERANSIM, an open source, software-based 5G RAN simulation tool that provides a comprehensive environment for emulating real-world 5G behaviors and supports research, testing, and validation of various deployment scenarios. It enables the deployment of multiple gNodeB instances, each configured with coverage parameters to mimic varied cellular environments \cite{7}. The system supports complete integration with the Next Generation Core (NGC) network by handling control plane signaling via the NG Application Protocol (NGAP) over the N2 interface and managing user plane traffic through the GTP-U protocol over the N3 interface, ensuring that the simulated network adheres closely to real-world 5G architectures \cite{24}. gNodeB instances are further designed to execute advanced radio resource management, dynamically allocating resources based on Quality of Service (QoS) requirements and facilitating seamless inter-gNodeB handovers to maintain continuous connectivity as UEs move across cells \cite{7}. 
In parallel, UERANSIM’s robust UE simulation capabilities enable the emulation of realistic user devices that support multiple PDU session types, including IPv4, IPv6, or dual-stack configurations, with each simulated UE operating with a complete 5G NAS protocol stack to accurately test processes such as network attachment, session establishment, and mobility management. This environment allows for fine-tuning of various network parameters, providing a versatile platform for in-depth analysis and optimization of both control and user plane interactions.

\subsubsection*{Activity Based Mobility Model}

To simulate realistic UE movement, we implement a custom activity-based mobility model within UERANSIM. This model captures human-like behavior by alternating between movement and dwell phases based on time-of-day and contextual preferences.

Each UE selects its next destination from a predefined set of activity locations, denoted as:
\[
\mathcal{L} = \{L_i\} = \left\{(x_i, y_i),\, \mu_i,\, \sigma_i \;\middle|\; i = 1,\dots,N \right\},
\]
where each location \(L_i\) is characterized by its coordinates \((x_i, y_i)\), and a dwell-time distribution modeled as \(\mathcal{N}(\mu_i, \sigma_i^2)\). Personal locations such as ``home'' and ``work'' are specified per the respective UE in its configuration file, while public destinations are shared, such as parks or coffee shops.
The day is divided into five time categories:
\[
t \in \{ \text{morning}, \text{lunch}, \text{afternoon}, \text{evening}, \text{night} \},
\]
Each category defines a probability distribution over activity types through a weight vector \(\mathbf{w}(t) = [w_1(t), w_2(t), \dots, w_K(t)]\). During morning hours, work-related destinations have higher weights, while recreational spots gain prominence in the evening.
When a UE is ready to select its next location:

\begin{enumerate}
    \item \textit{Activity Type Sampling:} A base type \(k\) is sampled according to \(\mathbf{w}(t)\), with a penalty applied to the current activity type \(k_0\) to reduce back-and-forth oscillations:
    \[
    w'_k(t) =
    \begin{cases}
    \epsilon \cdot w_k(t), & \text{if } k = k_0 \\
    w_k(t), & \text{otherwise}
    \end{cases}
    \]
    The adjusted weights are normalized to compute selection probabilities:
    \[
    P(k)=\frac{w'_k(t)}{\sum_j w'_j(t)}.
    \]
    
    \item \textit{Location and Movement:} A destination \(L_i\) is selected among those matching the sampled activity type. A base travel speed \(v_0\) is drawn from:
    \[
    v_0 \sim \mathcal{U}(v_{\min}, v_{\max}), \quad v = v_0 \cdot s(k, t),
    \]
    where \(s(k, t)\) adjusts speed to reflect contextual factors such as location type and congestion.
    
    \item \textit{Navigation and Dwell:} The UE navigates toward the selected location using the direction:
    \[
    \theta = \tan^{-1} \left( \frac{y_{\text{dest}} - y_{\text{current}}}{x_{\text{dest}} - x_{\text{current}}} \right),
    \]
    and moves at speed \(v\). Upon arrival, the UE remains stationary for a duration $\tau \sim \mathcal{N}(\mu_i, \sigma_i^2)$.
\end{enumerate}

This model produces temporally grounded and spatially meaningful movement traces that reflect real-world human behavior. As reported by previous experimentations in the field, the resulting traces drive UE interactions such as registration, deregistration, and handover events \cite{ernesto1}, which are logged and analyzed by our NWDAF implementation.

\section*{Data Generation and Analysis}\label{sec-6}

\begin{figure}
    \centering
    \includegraphics[width=0.5\linewidth]{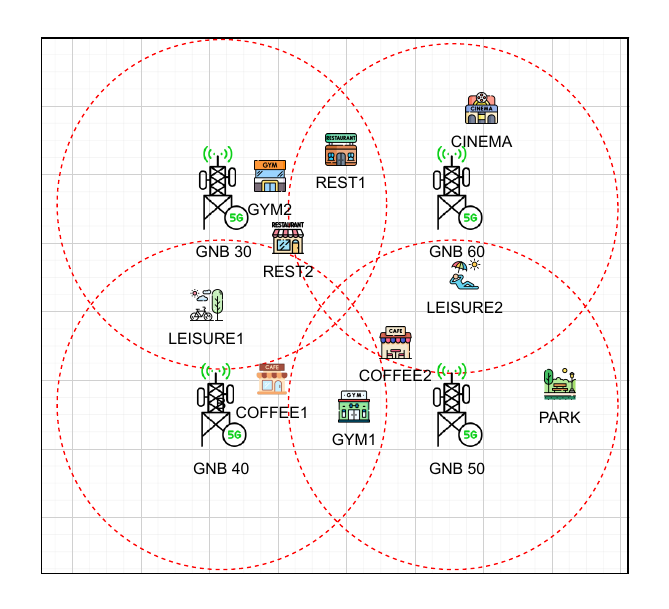}
    \caption{Mobility Region with GNB and Destination Locations}
    \label{fig:5}
\end{figure}
We evaluated the NWDAF-enhanced 5G network testbed over a two-week period to analyze the behavior of the mobility model and assess the operational capabilities of the NWDAF. The experiment included four virtual gNodeBs, each assigned a distinct tracking area code (TAC) and arranged in a square formation, as illustrated in Fig.~\ref{fig:5}.
Each gNodeB was configured with a simulated coverage radius of 120 units, corresponding to a signal strength threshold of -120 dBm (consistent with UERANSIM’s default configuration). Four virtual UEs were deployed across the network: three configured to dynamically attach and detach from the network, and one configured to maintain continuous connectivity throughout the experiment. Upon joining, each UE began from its designated home location. The simulation environment also included several public locations, such as coffee shops and parks, that UEs could visit, following the defined activity-based mobility model.

\begin{figure}[]
    \centering
    \begin{subfigure}[b]{0.5\textwidth}
        \centering
        \includegraphics[width=\textwidth]{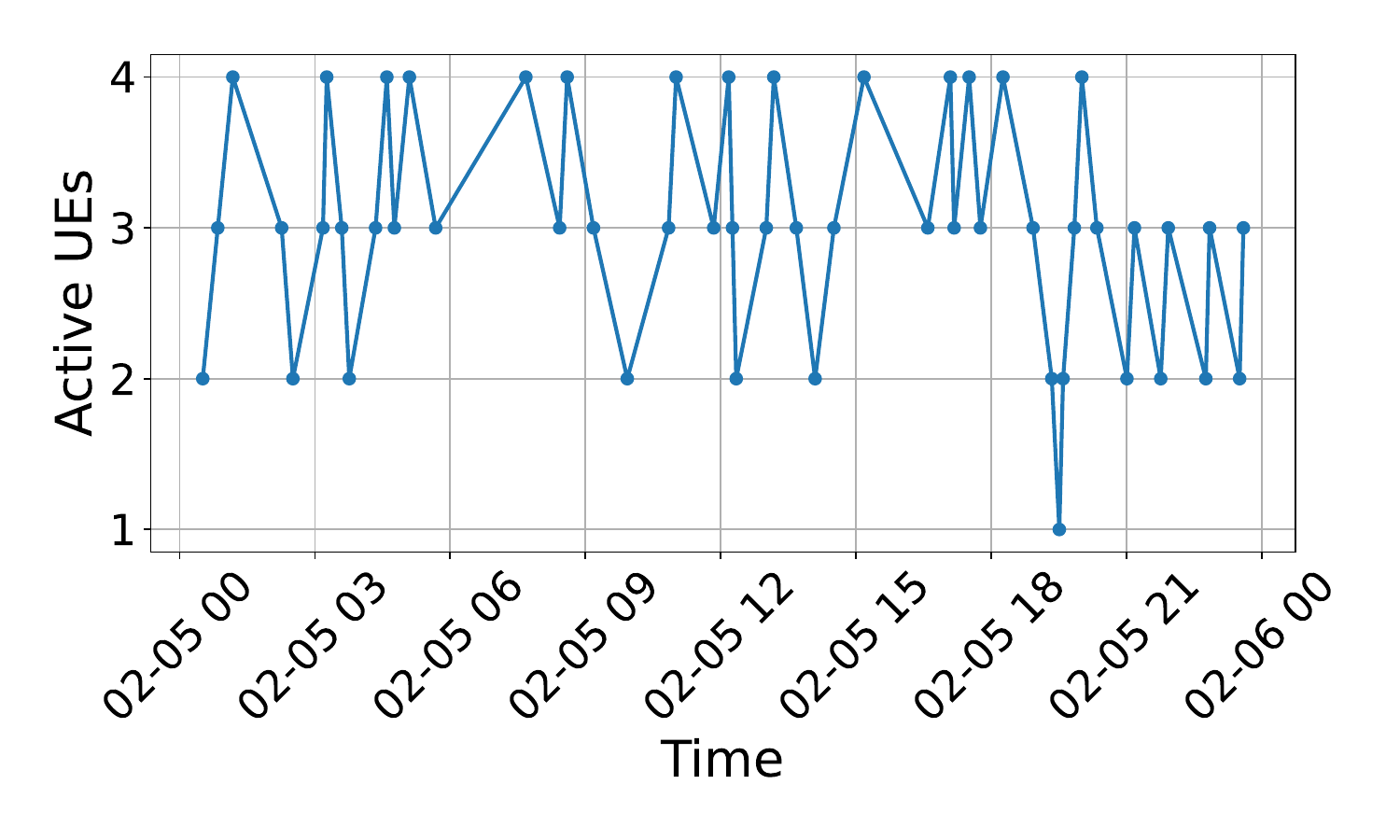}
        \caption{}
        \label{fig:active_ues}
    \end{subfigure}
    \hfill
    \begin{subfigure}[b]{0.5\textwidth}
        \centering
        \includegraphics[width=\textwidth]{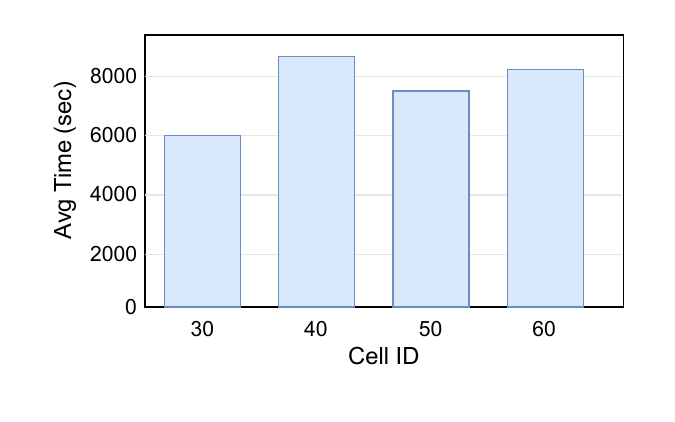}
        \caption{}
        \label{fig:avg_time}
    \end{subfigure}
    \hfill
    \begin{subfigure}[b]{0.5\textwidth}
        \centering
        \includegraphics[width=\textwidth]{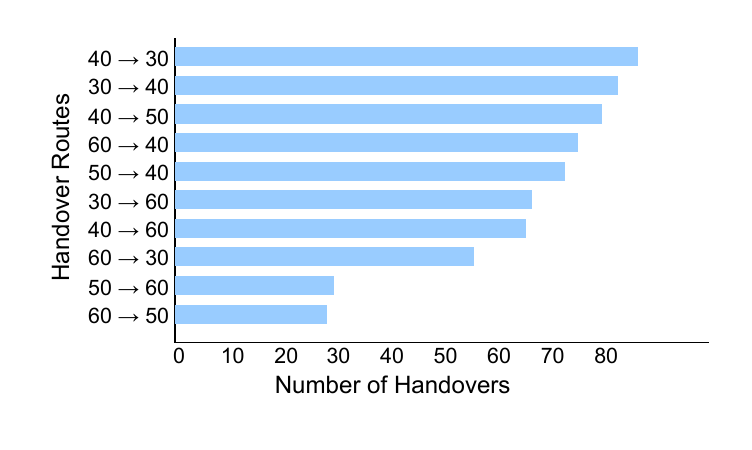}
        \caption{}
        \label{fig:handover_routes}
    \end{subfigure}
    \caption{Analysis of UE activity and handover behavior: (a) Number of active UEs throughout one day; (b) average time before handover; (c) handover routes}
    \label{fig:combined}
    \vspace{-2em}
\end{figure}
\subsection*{UE Registration, Monitoring, and Handover Behavior}

Throughout the test period, comprehensive data were collected about UE registration, state transitions, and handovers to evaluate the network’s ability to analyze dynamic user behavior. All registration and deregistration attempts were successful, with the core NFs reliably executing mobility management procedures. UEs exhibited distinct patterns in state durations, with the average time spent in the active state ranging from 100 to 102 minutes, while the inactive state typically lasted between 37 and 40 minutes.

As can be seen in Fig.~\ref{fig:active_ues}, Monitoring the number of active UEs over time showed clear temporal trends, which was expected due to the activity-based mobility model. The number of connected UEs varied throughout the day, with periods of high activity observed in the late morning around 11:00 and again between 14:00 and 17:00. While some hours had no active UEs, others reflected moderate load. The longest continuous activity session recorded for any UE was 9000 seconds (2.5 hours), demonstrating the system's ability to maintain stable connections over long durations.
Each handover event was logged with a timestamp, cell and device identifiers, and the TAC. These records indicate successful transitions across all configured cells in the testbed. One gNodeB emerged as a dominant traffic point, with significantly higher usage, due to its position with respect to the landmarks in the simulated topology. Handover frequency varied between UEs, with one device experiencing the most transitions, indicating either more aggressive movement or sensitivity to handover thresholds. Hourly trends showed that handovers were most common during mid-day and early evening hours, while early mornings remained relatively quiet. Certain cell pairs frequently exchanged UEs in both directions, pointing to either overlapping coverage zones or overly sensitive handover criteria. Dwell times within each cell also varied, with one particular cell showing the longest average residence time.
The NWDAF was actively engaged to collect analytics on UE mobility, registration events, and state transitions. Event subscription requests issued by the NWDAF were processed rapidly by the core network, with AMF and SMF acknowledging successful subscriptions in approximately 10 ms. Moreover, the NWDAF received and processed event notifications in approximately 109 ms after the events occurred. Performance monitoring revealed that the NWDAF operated with exceptional efficiency, utilizing merely 0.06\% of CPU resources and 0.17\% of system memory (27 MB RAM). All generated insights were logged and stored, and the resulting dataset is made available for future use. While our current setup simulates a controlled mobility pattern, we note that more realistic models can be incorporated by others to stress-test the framework further. This dataset and framework can serve as a foundation for extending NWDAF research and exploring richer scenarios.

\subsection*{Handover Prediction Model Evaluation
}
To evaluate the predictability of the UE handover events as a result of the activity-based mobility model, we employ four classification models on the UE location feature in the collected dataset, namely Random Forest, Gradient Boosting, K-Nearest Neighbors, and Decision Tree, on the dataset. The input features include the UE’s subscriber ID (SUPI), the IDs of the two most recently visited cells, time-of-day (the mobility model's time categories), the (x, y) coordinates of the most recent (current) cell, and the frequency with which the UE visited that cell at the given time period.

We split the dataset into training and testing sets corresponding to 70\% and 30\%, respectively. Gradient Boosting, with a learning rate of 0.05, maximum depth of 9, and 100 number of estimators, produced the highest accuracy (80.65\%), followed by Random Forest (80.24\%), while Decision Tree and K-Nearest Neighbors achieved an accuracy 80.11\% and 79.03\%, respectively. These results suggest that the collected data exhibits predictable patterns, making it a valuable starting point for further NWDAF-driven exploration and analytics. The models we compared are known to be robust against  unreliability or low quality of training data \cite{ernesto2}.

\subsection*{Intent Matching Evaluation
}
The LLM interface is designed to take natural language user queries, understand the intent, and act accordingly. Currently, our setup supports responding to queries related to subscription commands for the NWDAF or analytics requests based on the collected data. To evaluate the intent classification capabilities, we compared two approaches: (i) directly using LLMs (\textit{GPT-4o} and \textit{GPT-4o-mini}) to classify intents from text, and (ii) using embedding models (\textit{text-embedding-ada-002} and \textit{all-MiniLM-L6-v2}) that generate semantic vector representations of queries, followed by cosine similarity matching against a repository of predefined intents. Our experimental setup involved a dataset of 700 test prompts, systematically organized into seven categories (100 prompts each): \texttt{AMF Subscribe}, \texttt{AMF Unsubscribe}, \texttt{SMF Subscribe}, \texttt{SMF Unsubscribe}, \texttt{Active\_UEs}, \texttt{UE\_location\_report}, and \texttt{Registration\_State}. These categories encompass the current types of subscriptions and data collection the NWDAF is capable of. We also give the models a separate set of 1000 intent prompts, also organized into the same seven categories (150 prompts), for the models to use in matching the test prompts into their correct categories. Table \ref{tab:1} shows prompt examples from each category.
\begin{table}
    \centering
    
    \begin{tabular}{|l|l|p{0.15\textwidth}|}
    \hline
    Type & Category & Prompt example\\
    \hline
    Subscription management & AMF Subscribe & Set up NWDAF to receive AMF 
events\\
    \hline
    Subscription management & AMF Unsubscribe & Delete AMF event subscription \\
    \hline
    Subscription management & SMF Subscribe & I need NWDAF to monitor SMF \\
    \hline
    Subscription management & SMF Unsubscribe & Deactivate NWDAF's SMF 
monitoring capability\\
    \hline
    Analytics Query & Active UEs & How many UEs are currently active? \\
    \hline
    Analytics Query & UE location report & Show me the SUPIs that experienced handovers to NrCellId 9548800000 \\
    \hline
    Analytics Query & Registration state & What's the registration state trend over the course of the day? \\
    \hline
    \end{tabular}
    
    \caption{Prompt types with examples}
    \label{tab:1}
\end{table}
The primary objective of our testing was to assess the accuracy with which each model could classify these prompts into the correct intent categories. This classification is critical, as it reflects the model's ability to accurately interpret user intents, a fundamental requirement for any system that aims to automate responses in a dynamic network management environment. The performance of the \textit{GPT-4o} and \textit{GPT-4o-mini} models was compared against that of the \textit{text-embedding-ada-002} model and the \textit{all-MiniLM-L6-v2} model.

Our results demonstrated that the \textit{text-embedding-ada-002} model exhibited superior performance by correctly classifying 98.43\% of the prompts. Closely following was the \textit{all-MiniLM-L6-v2} model, which achieved an accuracy of 96.86\%. These results represent a significant improvement over the \textit{GPT-4o} and \textit{GPT-4o-mini} models, which achieved accuracies of 89.5\% and 55.1\% respectively. The high accuracy of the text-embedding model underscores its robustness in handling the complex semantics associated with 5G network management tasks. Unlike generative LLM-based classification, which can introduce hallucinations, the embedding approach relies on cosine similarity to match user queries against a predefined database of intents. This retrieval-based method ensures that the system selects from an existing, validated set of queries, thereby reducing ambiguity and enhancing the precision of automated intent recognition in network management.

\subsection*{LLM Analytics Prompt Answering Evaluation
}
To evaluate the LLM-based interface's ability to retrieve and present network insights, we conducted a query answering assessment. Unlike intent matching, which focuses on correctly categorizing user queries, query answering evaluates how well the system retrieves relevant, structured, and precise responses based on the collected network data. Below are five prompts that were used: 2 UE handover related, 2 Registration state related, and 1 Active UE count related. We present the operator's queries and the system’s responses, followed by an analysis of strengths and areas for improvement.

\begin{llmbox}

\textbf{1. Show me the handover pattern for SUPI 208930000000001}

\textbf{Response:}

\begin{quote}
Here is the handover pattern for the SUPI 208930000000001:

This handover pattern indicates how the device with SUPI 208930000000001 moves between different network cells and tracks over time, based on the data logged. The notable points from the pattern are:

Timestamps and Cell Connections:

\begin{itemize}
    \item The earliest connection in the dataset is at 2025-02-28 08:31:13 in cell 000000060, TAC 000001.
    \item The latest entry is at 2025-03-19 08:12:53, where it connects to cell 000000060, TAC 000001.
\end{itemize}
Cell and TAC Distribution:

\begin{itemize}
    \item The most frequently observed NrCellId for this SUPI is 000000060.
    \item The associated TACs show cell coordination changes, often alternating between TAC 000001 and TAC 000003.
\end{itemize}
Frequent Handover Locations:

The SUPI 208930000000001 has handed over between cells 000000030, 000000040, 000000050, and 000000060. These handovers indicate mobility within a consistent geographic area where these cells and tracking area codes are relevant.
Activity Pattern:

Activity is present throughout the day, showing consistent data logging and frequent handover\textbf{s} between cells. This suggests either continuous mobility or variable device movement where coverage changes require frequent transitioning.
\textit{If you need any more details or another specific timeframe highlighted within this dataset, let me know!}

\end{quote}

% \textbf{Analysis:}
\begin{tcolorbox}[colback=gray!5!white, colframe=gray!50!black, title=Analysis]
\begin{itemize}
    \item Correct Data Extraction. The system correctly identifies the earliest and latest handover events, frequent cells, and handover transitions.
    \item Unnecessary Filler Text. Some phrases such as \textit{"This handover pattern indicates how the device moves..."}, could be removed for conciseness.
\end{itemize}
\end{tcolorbox}
\end{llmbox}

\begin{llmbox}
\textbf{2. How many registration state changes occurred on 2025-03-18?}

\textbf{Response:}
\begin{quote}
To determine the number of registration state changes that occurred on 2025-03-18, let's filter the state change data for that specific date.

Analyzing the provided log:

\begin{itemize}
    \item 2025-03-18 10:55:45
    \item 2025-03-18 11:00:45
    \item 2025-03-18 11:00:45
    \item 2025-03-18 11:40:45
    \item 2025-03-18 11:45:45
\end{itemize}
Based on these entries, there were 5 registration state changes on 2025-03-18.

\end{quote}

% \textbf{Analysis:}
\begin{tcolorbox}[colback=gray!5!white, colframe=gray!50!black, title=Analysis]
\begin{itemize}
    \item Clear Answer with a Count. The system correctly counts and lists the timestamps.
    \item Could Include Context, a small note on whether this is above or below the average daily state changes would add value.
\end{itemize}
\end{tcolorbox}
\end{llmbox}

\begin{llmbox}
\textbf{3. Which SUPI has the most frequent state changes?}

\textbf{Response:}
\begin{quote}
To determine which SUPI has the most frequent state changes, we can calculate the total number of state changes for each SUPI. Based on the data provided, we can tally the occurrences of each SUPI's state transitions (from "active" to "inactive" or vice versa).

Here's a summary:

\begin{itemize}
    \item imsi-208930000000001 has gone through numerous state changes at different time intervals.
    \item imsi-208930000000002 has also experienced multiple state changes.
    \item imsi-208930000000003 appears to have a very high count of state changes.
    \item imsi-208930000000004 is mentioned at one timestamp, so it likely has fewer state changes.
\end{itemize}
However, based on the data provided, imsi-208930000000003 seems to consistently have very frequent state changes, making it the SUPI with the most dynamic behavior overall.

\end{quote}

% \textbf{Analysis:}
\begin{tcolorbox}[colback=gray!5!white, colframe=gray!50!black, title=Analysis]
\begin{itemize}
    \item Correct Answer, but lacks numbers. Instead of vague descriptions (\textit{"very high count"}), the system should return an exact number (e.g., \textit{"SUPI 208930000000003 had 18 state changes."}).
    \item Good Contextual Summary. The system correctly identifies the most dynamic UE.
\end{itemize}
\end{tcolorbox}
\end{llmbox}

\begin{llmbox}

\textbf{4. Show me the handover sequence for each SUPI in chronological order}

\textbf{Response:} To analyze the handover sequence for each SUPI in chronological order, let's break down the provided data to track the cell transitions over time: 

Here, the data is for four SUPIs: 

SUPI: 208930000000001 

2025-02-28 08:31:13: NrCellId 000000060 

2025-03-03 09:03:37: NrCellId 000000030 

2025-03-05 16:17:58: NrCellId 000000060 

2025-03-06 02:21:34: NrCellId 000000040 

2025-03-07 09:54:46: NrCellId 000000040 

2025-03-09 14:43:44: NrCellId 000000050 

2025-03-11 00:43:19: NrCellId 000000060 

2025-03-14 10:06:05: NrCellId 000000060 

... 

SUPI: 208930000000002 

2025-02-28 08:31:13: NrCellId 000000040 

2025-03-03 09:03:37: NrCellId 000000030 

2025-03-05 16:17:51: NrCellId 000000040 

2025-03-06 08:14:23: NrCellId 000000060 

2025-03-07 09:54:46: NrCellId 000000040 

2025-03-09 14:43:44: NrCellId 000000050 

2025-03-11 00:43:19: NrCellId 000000060 

2025-03-14 10:06:05: NrCellId 000000060 

... 

SUPI: 208930000000003 

2025-02-28 08:30:59: NrCellId 000000040 

2025-02-28 08:31:13: NrCellId 000000060 

2025-02-28 09:16:29: NrCellId 000000040 

2025-02-28 10:26:59: NrCellId 000000030 

2025-03-03 09:03:40: NrCellId 000000030 

2025-03-05 16:17:58: NrCellId 000000040 

2025-03-05 18:42:08: NrCellId 000000060 

2025-03-05 23:56:07: NrCellId 000000040 

... 

SUPI: 208930000000004 

2025-02-28 08:30:59: NrCellId 000000040 

2025-02-28 08:31:13: NrCellId 000000060 

2025-02-28 08:49:11: NrCellId 000000050 

2025-02-28 09:16:29: NrCellId 000000040 

2025-02-28 10:26:59: NrCellId 000000030 

2025-03-03 09:03:40: NrCellId 000000030 

2025-03-05 16:17:51: NrCellId 000000040 

2025-03-05 18:42:08: NrCellId 000000060 

... 

This sequence details how each SUPI transitions between different cell IDs (NrCellId) over time. The unique identifier NrCellId and time indicate the moment a mobile device was connected or handed over to a particular cell, and tac (Tracking Area Code) represents different geographical tracking areas. For a more in-depth analysis, you might want to extract complete sequences and monitor cell transitions for network optimization.

% \textbf{Analysis:}
\begin{tcolorbox}[colback=gray!5!white, colframe=gray!50!black, title=Analysis]
\begin{itemize}
    \item Great Detailed Response. The system returns the full handover sequence per SUPI.
    \item Could Use a Timeline Format, presenting results as a chronological table or graph would improve readability.
\end{itemize}
\end{tcolorbox}
\end{llmbox}

\begin{llmbox}

\textbf{5. What’s the active UE pattern?}

\textbf{Response:}

The pattern of active User Equipment (UE) over the period provided can be inferred from the data. Here is a summary:

Time Frame:

\begin{itemize}
    \item The data encompasses timestamps from early March 2025 up to March 18, 2025.
    \item The activity is distributed over morning, afternoon, evening, and night timestamps.
\end{itemize}
Active UE Values:

The number of active UEs recorded at each timestamp varies between 1 and 4.
General Observations:

\begin{itemize}
    \item Higher UE activity (values of 3 and 4) suggests peak utilization periods, often during daytime and early evening.
    \item Lower UE activity (values of 1 and 2) tends to occur late at night and early in the morning, indicating off-peak periods.
\end{itemize}
Duration Between Changes:

Some intervals have long durations where the active UE count remains stable, while others show frequent fluctuations.
Irregular Activity:

Specific time periods, such as March 3 and March 14-18, may show less consistent data, possibly due to system maintenance or external influences.

  % \textbf{Analysis:}
\begin{tcolorbox}[colback=gray!5!white, colframe=gray!50!black, title=Analysis]
\begin{itemize}
    \item The response correctly identifies variations in active UE counts, but the explanation includes unnecessary speculation about external influences. In the context of our controlled testbed, it is more appropriate to concisely report the observed numerical patterns without attributing them to unverified causes. However, in real-world networks, such speculation can be valuable for root-cause analysis and operational troubleshooting.
\end{itemize}
\end{tcolorbox}
\end{llmbox}

\section*{Conclusion}\label{sec-7}

This paper presents the implementation of a software-based 5G network testbed that combines Free5GC, extended with a fully functional NWDAF enhanced by an LLM-powered interface, and UERANSIM, extended with a custom mobility model. We implement event subscription mechanisms for the NWDAF and automate network analysis through a RAG-based LLM-powered NWDAF. Through experimentation, we validated the system’s capability to capture network events, facilitate event-driven analytics, and predict UE behavior based on historical trends. The integration of an LLM-based interface significantly improves the usability of the NWDAF, allowing operators to interact with the system using natural language.

A key direction for future development lies in enhancing the LLM interface to support more advanced automation and policy enforcement within the 5G core network. Rather than limiting the LLM to intent classification and basic analytics queries, future iterations can enable the system to autonomously request supplementary data from relevant NFs when needed to fulfill a user’s prompt. This will allow for more dynamic and context-aware responses. Moreover, future work can focus on reducing the cost of leveraging LLMs for these tasks, for example by employing specialized small language models or model compression techniques such as quantization and pruning. Additionally, future work can focus on generating and invoking new NWDAF service functions based on user intents, bridging the gap between human-friendly commands and low-level network control tasks. Such capabilities would move the system closer to a fully intelligent and self-operating network.

\section*{Data Availability}
The source code and datasets generated during the current study are available in the github repository, https://github.com/HenokDanielbfg/testbed.

% \section*{Statements and Declarations}
% \begin{itemize}
%     \item \textbf{Funding}: The authors did not receive support from any organization for the submitted work.
%     \item \textbf{Conflict of Interest}: On behalf of all authors, the corresponding author states that there is no conflict of interest..
% \end{itemize}

% \bibliographystyle{sn-mathphys-num}
% \bibliographystyle{nature}
\bibliography{references}

% \section*{Introduction}

% The Introduction section, of referenced text\cite{Figueredo:2009dg} expands on the background of the work (some overlap with the Abstract is acceptable). The introduction should not include subheadings.

% \section*{Results}

% Up to three levels of \textbf{subheading} are permitted. Subheadings should not be numbered.

% \subsection*{Subsection}

% Example text under a subsection. Bulleted lists may be used where appropriate, e.g.

% \begin{itemize}
% \item First item
% \item Second item
% \end{itemize}

% \subsubsection*{Third-level section}
 
% Topical subheadings are allowed.

% \section*{Discussion}

% The Discussion should be succinct and must not contain subheadings.

% \section*{Methods}

% Topical subheadings are allowed. Authors must ensure that their Methods section includes adequate experimental and characterization data necessary for others in the field to reproduce their work.

% \bibliography{sample}

% \noindent LaTeX formats citations and references automatically using the bibliography records in your .bib file, which you can edit via the project menu. Use the cite command for an inline citation, e.g.  \cite{Hao:gidmaps:2014}.

% For data citations of datasets uploaded to e.g. \emph{figshare}, please use the \verb|howpublished| option in the bib entry to specify the platform and the link, as in the \verb|Hao:gidmaps:2014| example in the sample bibliography file.

\section*{Author contributions statement}

% Must include all authors, identified by initials, for example:
% A.A. conceived the experiment(s),  A.A. and B.A. conducted the experiment(s), C.A. and D.A. analysed the results.  All authors reviewed the manuscript. 
H.D. implemented the system and executed the experiments. O.A. conceptualized the study and served as the project administrator. C.L. and J.L. provided critical methodological review and contributed essential expertise to the interpretation of the results and discussion. E.D. performed the technical review of the system implementation and provided critical intellectual revision of the manuscript. All authors approved the final version.

\section*{Additional information}
\begin{itemize}
    \item \textbf{Competing Interest}: The authors declare no competing interests.
    \item \textbf{Funding Statement}: No funding received.
\end{itemize}
% To include, in this order: \textbf{Accession codes} (where applicable); \textbf{Competing interests} (mandatory statement). 

The corresponding author is responsible for submitting a \href{http://www.nature.com/srep/policies/index.html#competing}{competing interests statement} on behalf of all authors of the paper. This statement must be included in the submitted article file.

\section*{Appendix}
% \section{Data Collection Samples}
The following tables present representative samples of the data collected in real-time from AMF and SMF via the NWDAF's event subscription framework. This data forms the foundation of our analytics and machine learning modules.

The NWDAF subscribes to event notifications from both the AMF and the SMF. As subscribed events occur, such as UE registration, PDU session changes, or handovers, the respective NF sends an event report to the NWDAF using HTTP callbacks. These event reports are then parsed, formatted, and exposed as Prometheus metrics by the NWDAF's internal consumer module. This allows real-time monitoring, querying, and visualization using standard observability tools.

In this appendix, we include sample records from three key Prometheus metrics, each of which corresponds to one of the tables below:
\begin{itemize}
    \item Table \ref{tab:active_ues} - \verb|Active_UEs|: Tracks the number of user devices actively connected to the network at each timestamp, along with the duration they remain in that state. This metric supports session management analysis and capacity planning.
    \item Table \ref{tab:handovers} - \verb|UE_location_report|: Captures granular handover data, including the UE identifier (SUPI), the current serving cell (NrCellId), and its location area (TAC). This metric underpins mobility behavior tracking and handover prediction modeling.
    \item Table \ref{tab:registration state} - \verb|UE_registration_state|: Logs registration state transitions (active/inactive) per UE, alongside timestamps and state durations. This data is vital for evaluating user engagement patterns and optimizing registration state transitions.
\end{itemize}
The complete datasets for these metrics are available in the accompanying GitHub repository.

\begin{table}[]
    \centering
    \begin{tabular}{|c|c|c|}
        \hline
        Timestamp & Active UE & Duration \\
        \hline
        2025-03-10 11:30:45 & 3 & 0 days 00:35:00 \\
        2025-03-10 12:05:45 & 4 & 0 days 00:15:00 \\
        2025-03-10 12:20:45 & 3 & 0 days 00:35:00 \\
        2025-03-10 12:55:45 & 2 & 0 days 00:10:00 \\
        2025-03-10 13:05:45 & 3 & 0 days 00:10:00 \\
        2025-03-10 13:15:45 & 4 & 0 days 00:10:00 \\
        2025-03-10 13:25:45 & 3 & 0 days 00:40:00 \\
        2025-03-10 14:05:45 & 4 & 0 days 01:25:00 \\
        2025-03-10 15:30:45 & 3 & 0 days 00:15:00 \\
        2025-03-10 15:45:45 & 2 & 0 days 00:30:00 \\
        \hline
    \end{tabular}
    \caption{Active UEs in dataframe format}
    \label{tab:active_ues}

    \begin{tabular}{|c|c|c|c|}
        \hline
        Time & NrCellId & SUPI & TAC \\
        \hline
        2025-03-05 16:17:51 & 000000040 & 208930000000004 & 000003 \\
        2025-03-05 16:17:58 & 000000040 & 208930000000003 & 000003 \\
        2025-03-05 16:17:58 & 000000040 & 208930000000002 & 000003 \\
        2025-03-05 16:17:58 & 000000060 & 208930000000001 & 000001 \\
        2025-03-05 16:20:53 & 000000030 & 208930000000001 & 000004 \\
        2025-03-05 16:22:28 & 000000060 & 208930000000002 & 000001 \\
        2025-03-05 16:29:58 & 000000050 & 208930000000003 & 000002 \\
        2025-03-05 17:33:50 & 000000050 & 208930000000002 & 000002 \\
        2025-03-05 18:20:24 & 000000040 & 208930000000003 & 000003 \\
        2025-03-05 18:42:08 & 000000060 & 208930000000001 & 000001 \\
        \hline
    \end{tabular}
    \caption{UE location report in dataframe format}
    \label{tab:handovers}

    \begin{tabular}{|c|c|c|c|}
        \hline
        \textbf{Timestamp}       & \textbf{SUPI}               & \textbf{State Description} & \textbf{Duration (mins)} \\
        \hline
        2025-03-10 11:30:45  & 208930000000001  & inactive          & 10.0                 \\
        2025-03-10 11:30:45  & 208930000000003  & active            & 50.0                 \\
        2025-03-10 11:30:45  & 208930000000002  & active            & 10.0                 \\
        2025-03-10 11:30:45  & 208930000000004  & active            & NaN                  \\
        2025-03-10 11:40:45  & 208930000000001  & active            & 105.0                \\
        2025-03-10 11:40:45  & 208930000000002  & inactive          & 25.0                 \\
        2025-03-10 12:05:45  & 208930000000002  & active            & 50.0                 \\
        2025-03-10 12:20:45  & 208930000000003  & inactive          & 45.0                 \\
        2025-03-10 12:55:45  & 208930000000002  & inactive          & 20.0                 \\
        2025-03-10 13:05:45  & 208930000000003  & active            & 160.0                \\
        \hline
    \end{tabular}
    \caption{UE Registration state in dataframe format}
    \label{tab:registration state}
\end{table}

% Figures and tables can be referenced in LaTeX using the ref command, e.g. Figure \ref{fig:stream} and Table \ref{tab:example}.

\end{document}